\documentclass[preprint]{aastex63}
\usepackage{xcolor}

\newcommand{\gray}{$\gamma$-ray\ } \newcommand{\grays}{$\gamma$-rays\ }

\received{April 19, 2020}
\revised{September 19, 2020}
\accepted{January 7, 2021 }
\published{2021 in The Astrophysical Journal, Volume 909, page 52; 
DOI: 10.3847/1538-4357/abda3f}

\shorttitle{LyC and \gray absorption}
\shortauthors{}

\begin{document}

\title{Gamma-Ray Absorption By The Cosmic Lyman Continuum from Star-forming Galaxies}

\correspondingauthor{Mathew A. Malkan}
\email{malkan@astro.ucla.edu}

\author{Matthew A. Malkan}
\affiliation{Department of Physics and Astronomy\\ 
University of California, Los Angeles \\ 
Los Angeles, CA 90095-1547}

\author{Sean T. Scully}
\affiliation{Department of Physics and Astronomy\\
James Madison University\\
Harrisonburg, VA 22807}


\author{Floyd W. Stecker}
\affiliation{Astrophysics Science Division\\
NASA/Goddard Space Flight Center\\
Greenbelt, MD 20771}
\affiliation{Department of Physics and Astronomy\\ 
University of California, Los Angeles \\ 
Los Angeles, CA 90095-1547}



\begin{abstract}

Motivated by the discovery of the ultra-strong emission line starburst galaxies (EELGs) 
known as ``green pea galaxies",
we consider here their contribution to the intergalactic flux of ionizing UV at high redshifts.
Most galaxies that have been observed show a precipitous drop in their flux blueward of the Lyman limit.
However, recent observations of EELGs have discovered that
many more Lyman continuum photons escape from them into intergalactic space than was previously
suspected. We calculate their contribution to  the extragalactic background light (EBL).
We also calculate the effect of these photons on the absorption of high energy \grays.
For the more distant \gray sources, particularly
at $z \ge 3$, the intergalactic opacity above a few GeV is significantly higher than previous estimates which ignored
the Lyman continuum photons.
We calculate the results of this increased opacity on observed \gray spectra, which produces a high-energy turnover starting at lower 
energies than previously thought, and a gradual spectral steepening that may also be observable. 
\end{abstract}

\vspace{2.5cm}

\section{Introduction} \label{sec:intro}

Within the past decade several empirically based calculations of the Extragalactic Background Light (EBL) have been published 
\citep{stecker2012,stecker2016,helgason2012,scully2014}. These calculations have become increasingly precise, 
since they are now constrained by direct integration of extensive Luminosity Functions (LFs) of galaxies across most of cosmic time. 
In the latest determination, the EBL from the UV through to the FIR from $z = 0$ to $z = 8$ has been derived completely from observational data, 
without any modeling  assumptions about how the galaxies evolve with cosmic time \citep{stecker2016}. 
The EBL is of intrinsic interest as it encodes information about the star formation rate history of the universe. 
It is also fundamentally important because it determines the \gray opacity of
intergalactic space. 
The opacity is caused by the annihilation of \grays by $e^{+}$e$^{-}$ pair production with these low energy photons. 
As first suggested by \citet{stecker1992}, \gray observations from high redshift sources such as blazars (and later \gray bursts) 
can therefore constrain the EBL. 

For many years, the flux of ionizing photons leaking from galaxies could hardly be detected by direct observations. 
In addition, it could not be reliably modeled theoretically, owing to uncertainties in both the intrinsic FUV spectra of O stars and the fraction of FUV flux escaping H I absorption in their host galaxies. 
However, recent observations indicate that more LyC photons at high redshifts leak from galaxies into intergalactic space than was previously thought. 

While there has been some theoretical modeling that has included of the effect of LyC radiation on \gray opacity 
\citep{salamon1998,inoue2013}, 
most calculations of  $\gamma$-ray opacity have not accounted for
the possible amount of intergalactic far-UV (FUV) photons at wavelengths below that of Ly$\alpha$, and especially the 
Lyman continuum (LyC) photons with wavelengths $\lambda< $ 91.2 nm. 

More recently there has been detailed modeling of the EBL that included ionizing UV  eg., \citep{khaire2019,puchwein2019}.
Such ionizing photons, whether they be from galaxies or from quasars, can contribute to the reionization of the intergalactic medium. 
This process also plays an important role in determining the past and present structure of the Universe.

The estimated increase in the flux of LyC photons, in turn, increases the high-redshift opacity estimates for \grays at energies below 150 GeV. The cross-section for $e^{+}$e$^{-}$ pair production has a broad peak at a value for the square of the center-of- momentum energy of $\sim$(1 MeV)$^2 \simeq$ 1 eV TeV. 
Thus, a 15 eV LyC photon will most efficiently interact with a 67 GeV \gray. 
This \gray would have a present observed energy of $ \sim 67/(1+z)$ GeV. 
Thus, \grays emitted at a redshift of 3 interacting with 15 eV ionizing photons would be absorbed within the  energy range for detectability by the {\it Fermi}-LAT space telescope. 
Our new opacity predictions can thus be tested against {\it Fermi} observations of sources at $2 < z < 4.3$. Conversely, these {\it Fermi} data directly probe the ionizing EBL at high redshifts, providing new information on cosmic re-ionization.
 
Observations of most galaxies show a precipitous drop-off in their spectra beyond the Lyman limit. (E$ > $13.6 eV).
Extensive attempts to detect rest-frame LyC photons escaping from galaxies have proven extremely difficult, often requiring sensitive
space observations at the shortest UV wavelengths, of $0.3 < z < 2$ galaxies.

One early example was deep imaging observations using the Space Telescope Imaging Spectrograph on the Hubble Space Telescope 
(HST/STIS) \citep{malkan2003}. 
Imaging of 11 massive star-forming galaxies in the redshift range $1.1 < z < 1.4$ showed that the fractions of LyC photons emitted 
by high-mass stars that escape the host galaxy ($f_{esc}$),  were less than 1\% - 6\% of those predicted 
from their observed near-UV continuum radiation. 
Subsequently, HST observations were used to study a larger sample of 600 $z \sim 1$ galaxies having an average flux fraction $f_{esc} <$ 2\%, of the fluxes predicted for the near-UV continuum radiation \citep{rutokowski2016}. 
Other searches for LyC photons in some closer, brighter galaxies were performed using the FUSE (Far Ultraviolet Space Explorer) telescope. 
FUSE only detected LyC in two dwarf starbursts ($f_{esc}$ = 3\% - 5\%) within its FUV sensitivity range \citep{leitet2013}.  
At this impasse it seemed that only the much rarer quasars were contributing significantly to the flux of ionizing photons into the EBL. 

Searches for leakage of ionizing photons have also been made at higher redshifts, where the Lyman limit is shifted to visible wavelengths. 
A recent spectroscopic survey of 124 galaxies at redshift $z \sim 3$ has made weak detections of 15 of them below the Lyman limit, 
specifically those with the strongest Lyman-$\alpha$ emission lines \citep{steidel2018}.  

These LyC measurements at higher redshifts ($z > 3$) are more difficult because numerous intervening HI clouds in the IGM can absorb LyC emission from high redshift galaxies. 
Since the amount of IGM absorption varies from one sight-line to another, interpretations from emitted to observed radiation are difficult to make.  
Another problem is that some claimed detections of LyC turned out to be artifacts of emission at longer wavelengths, from contaminating low-redshift galaxies that had not been recognized close to the line-of-sight \citep{vanzella2015}.
However, several recent discoveries of substantial fluxes of ionizing photons from certain types of starburst galaxies is changing this situation.

\section{Strong Emission-Line Galaxies} \label{sec:gp}

Several recent detections of substantial fluxes of ionizing photons escaping from galaxies are transforming the observational situation outlined in the previous section.
Recent searches for escaping Lyman continuum photons have particularly focused on compact galaxies with
extreme high star formation rates. 
The HST Cosmic Origins Spectrograph measured a LyC escape fraction of 21\% in a starburst galaxy at $z = 0.23$  \citep{borathkur2014}, and two other dwarf starbursts with relative escape fractions of 21\% \citep{leitherer2016}. 
(The relative escape fraction is defined as the fraction of photons observed shortward of the Lyman limit compared with the number 
predicted based on the observed flux just longward of the limit.) 
Several recent papers present ``gold-standard" spectroscopic detections of LyC photons escaping from dwarf starburst galaxies at $z = 0.3$ \citep{borathkur2014}. Five of these have $f_{esc}$ of 10\% or greater \citep{izotov2016}.

These and other searches suggest that the most likely sources of escaping LyC emission are the most extreme young starbursts that concentrate their energy release in compact regions, thus opening ionized holes through their normally opaque HI gas. 
These starbursts are very rare in the low-$z$ universe, but were widespread at redshifts $6 < z < 9$, when they re-ionized the Universe. 
The best indicators of this phenomenon appear to be extremely strong [OIII] emission (EW([OIII]) = 200--1000 \AA) and extremely high ratios of [OIII]/[OII] $> 5$.   
There is a good correspondence of high [OIII]/[OII] with strong LyC escape, with $f_{esc}$ = 6\% to 13\% in 5 observed galaxies \citep{izotov2016}. 

These ultra-strong [OIII]-emitters belong in the new class of so-called "green pea" galaxies, noticed by citizen-scientists when they were examining color galaxy images from the Sloan digital sky survey (SDSS) \citep{cardamone2009}. 
These are compact, metal-poor dwarf galaxies undergoing extremely strong starbursts. 
As a result of their unusual large populations of very young O stars, so much of their energy emerges in the [OIII] 501+496 nm emission line doublet (EW([OIII] $\ge 200$\AA), 
that it impacts the overall {\it broadband colors}. 
This gives them their distinctive green appearance. 
We operationally define green pea galaxies to be extreme emission-line galaxies (EELGs) having [OIII] equivalent widths of 200 \AA ~or greater \citep{atek2014}. 
Because of their very high ionization, their [OIII]/[OII] 501/373 nm emission line ratios are
extremely high, i.e., greater than 4. 

\subsection{ Redshift Evolution of Lyman Escape Fraction}

Mounting observational evidence over the last two decades shows that hot stars in high-redshift galaxies systematically produce
more extreme-UV photons than those observed locally.
Emission lines from most current ($z = 0$) star-forming regions are dominated by singly ionized gas, produced by LyC photons slightly above 13.6 eV. 
However, doubly ionizing oxygen requires ionizing EUV photons with energies of above 35 eV.  
The HII regions in high-redshift galaxies have very highly ionized gas that is not described by the photoionization models developed for O stars in local galaxies \citep[e.g.][]{malkan1996}. 
The reasons that the O stars in the young universe have much harder
spectra than those observed locally are under active debate. 
Possible explanations range from low metallicities to rapid rotation in binary stars.  Regardless of the explanation,
it is an empirical fact that a growing number of dwarf galaxies with extremely high [OIII]/[OII] ratios and low 
abundances are now also known to emit LyC continuum \citep{izotov2018a,izotov2018b}.

At high redshifts, detections of escaping Lyman limit photons are still limited to a relative handful of individual galaxies, 
and to statistical detections in co-added stacks of observations of large samples. 
An example of the former is the measurement of ionizing radiation in a Ly$\alpha$-emitting galaxy at  $z = 2.37$ that
has been observed through
gravitational  lensing.  \citet{rivera2019} detected Lyman limit photons from twelve magnified individual clumps  of a
gravitationally lensed arc.  These show a range of relative escape fractions in various sight-lines to the high-ionization dwarf galaxy 
separated by several hundred parsecs, from 19\% to 64\%.

Recent stacking analyses of the Lyman continuum emission of 111 luminous galaxies at $z \sim 3$ in the GOODS fields \citep{smith2020}
and with 201 galaxies at $z \sim 4$ \citep{marchi2018}
are consistent with the trends we have identified:

1) The escape fractions of Lyman limit photons 
for the more massive galaxies   
are low (less than 5\%); 

2) The escape fractions of Lyman limit photons from galaxies are generally low for $z \le 3$; and 

3) The relative escape fractions are $ \sim 8\% -10\% $ in small galaxies, or those with
strong Ly$\alpha$ emission lines.

This limited sample is sufficient to reveal clear systematic trends that we will use. 
The good correlations of $f_{esc}$ with [OIII] and Ly$\alpha$ emission line strengths provide
us with a quantitative statistical basis to predict the total escaping LyC backgrounds.
The empirical correlation between [OIII] strength and LyC escape fraction, $f_{esc}$, implies  
substantial amounts of LyC photons leak into the IGM at high redshifts. 

\vspace{3.5cm}

\section{The Cosmic Lyman Continuum Background} \label{sec:ef}

\subsection{ Redshift Evolution of Extreme Emission Line Galaxies} 

Because of the general aging of galaxies over cosmic time, it has long been known 
that the higher the galaxy's redshift, the stronger are its emission lines produced by ionizing photons from young stars.
An extension of this trend is that the relative number of green pea galaxies increases with redshift. 
At earlier cosmic times, the green pea phenomenon is observed in more massive galaxies, an effect referred to as cosmic downsizing. 
Observational studies have recently begun to measure this  very strong redshift evolution
of green pea galaxies, all the way back to the epoch of cosmic reionization \citep{li2018}. 

The first discovered green pea galaxies appeared to be flukes in the local universe. 
Out of $\sim 10^6$ spectra in the local volume covered by the SDSS, only a few dozen had been verified. 
They are
very rare at redshifts $z \sim 0.3$ \citep{brunker2020}.
However, many more green pea galaxies were subsequently discovered at 
higher redshifts up to 1 using sensitive narrow-band imaging surveys \citep{ly2007}. 
Even higher densities of them have been found from slitless GRISM spectroscopy surveys at redshifts between 1 and 2 
using the Hubble Space Telescope \citep{atek2014}. 
More recently \citet{li2018} conducted a wide area photometric search for green pea galaxies using SDSS imaging. 
The green pea galaxy population that they discovered at $z \sim 0.5$ extends up to several times higher [OIII] luminosities 
than the low redshift ($z \sim 0.1$) green pea galaxies. 
This evolution continues to higher redshifts. The ``knee' in the [OIII] line luminosity function brightens by 
2.5 times from z=0.41 to z=0.83 \citep{ly2007}.
At $z \sim 1.5$ strong [OIII] emitters are ten times more numerous than at $z \sim 0.1$ \citep{maseda2018}.

Decisive statistics at $z \sim 3$ come from a recent analysis of 1294 galaxies in the Subaru Deep Field.  
Their overwhelmingly strongest emission line, viz. the [OIII] 501/496 nm doublet, is so strong that it dominates their entire rest-frame optical spectra (average rest-frame equivalent width of $\sim$ 1000 \AA). 
Stacked SEDs of thousands of U-dropout galaxies at $z$ = 3.5 show that as the stellar mass of the {\it average} Lyman break galaxy drops from $10^9$ to $10^8 M_{\odot}$, the rest equivalent width of ([OIII]501+496) rises from 400 to $> 1000$ \AA\  \citep{malkan2017}.
Such extreme emission line galaxies appear to be even more universal at yet higher redshifts, as indicated by their IRAC-band excess flux due to extremely strong [OIII] emission \citep{smit2015}.

The dominance of green pea galaxies is
further amplified because the UV luminosity function where green pea galaxies are the most frequent becomes extremely steep at higher redshifts. 
In fact, the faint-end slope of approximately $-2$  almost causes the integrated total UV light to diverge \citep{bouwens2015, malkan2017}. 
{\it Thus at $ z > 3$, the presently rare green pea galaxies were actually the most numerous galaxy type.}

\subsection{ LyC Escape Fraction from Green Pea Galaxies as a function of redshift}

Despite the observational challenges, quite a few estimates of LyC escape fraction now exist  for individual galaxies  for redshifts of $z \le 0.5 $, 
some of which exceed ten percent. 
For example, HST/COS has detected a 21\% escape fraction in a  starburst galaxy at z=0.23 \citep{borathkur2014}, 
and two other dwarf starbursts with relative escape fractions of 21\% \citep{leitherer2016}.
While there are outliers, the overall trend found is an increase in escape fraction with redshift.  In support of this trend, 
in the $z < 0.5$ universe, galaxies with large $f_{esc}$ are primarily of the compact green-pea type, characterized by their low metallicities and 
very strong nebular emission-lines \citep[e.g.][]{izotov2016,vanzella2020}. At $z \sim 0.5$, a very rough observational estimate 
might be that half of the green pea
galaxies have large (20\%) Lyman continuum relative escape fractions  \citep{malkan2020}. 

Green-pea galaxies have star-formation rates in excess of what is measured in local star-forming galaxies,
but may be comparable to those at higher redshifts during the epoch of reionization. 
\
At $z$ = 3.2 one such green pea galaxy with has been detected with (EW([OIII]) $\sim$ 1000 \AA) and with its LyC emission having an escape fraction of 64\% \citep{debarros2016}.  
A recent HST imaging study of the Lyman limits of 61 galaxies at $z = 3.1$ \citep{fletcher2019} finds that 12 of them have significant fluxes of escaping LyC photons with $f_{esc} \ge$ 15\% \citep{fletcher2019}. 
Four out of eight of these detected Lyman leakers have strong [OIII] lines with rest-frame 501+496 nm lines having equivalent widths of 100--1000 \AA. The [OIII]/[OII] ratios, where measured, were in the range 6 - 10, confirming their green pea classification.
Further supporting evidence that substantial fluxes of LyC photons escape from the fainter but more typical galaxies at $z \ge 3$ comes from 
spectroscopy of the galaxies hosting gamma ray bursts \citep{vielfaure2020}. These galaxies have relative LyC escape fractions of 8\% to 45\%.
 
At higher redshifts, the observed green pea type EELG starbursts with overwhelming [OIII] emission and strong leakage of FUV and EUV photons is no longer limited to dwarf galaxies, but becomes the norm for most galaxies.

A Lyman continuum escape fraction and EUV spectrum that varies systematically 
with galaxy luminosity and strongly with redshift requires a new parameter to make the
EBL calculation more realistic. Therefore, we will constrain our model calculations using three 
statistical observations:

1) The average equivalent width of [OIII] is now determined by taking averages of hundreds to
thousands (where available) of spectral energy distributions 
This empirically constrains the proportion of EELGs at $z = 1.5$, 2.0, 3.0, and 5.0 and 6.0 to 7.0.

2) The escape of LyC photons from galaxies is well correlated with the escape of Ly$
\alpha$ photons \citep{verhamme2015}.  
A possible physical explanation for this correlation is that both quantities are
driven by increasing porosity of a galaxy's ISM \citep{gazagnes2020}. 
Our LyC escape fractions should therefore also follow the observed rapidly increasing average equivalent width of Ly$\alpha$, 
which approximately doubles from $z = 4 $ to $z = 6 $ \citep{treu2013}.

3) The integrated LyC photon flux should equal the amount required to reionize the
intergalactic medium. 
Calculations indicate that the average Lyman escape fraction for all galaxies must be $\sim $ 20\% in order to reionize the Universe \citep{robertson2015,khaire2016,wise2014}. 
The discovery of a sudden drop in the number of Ly$\alpha$ emitters (LAEs) between $z = 5.7$ and $z = 6.6$ appears to be caused by extrinsic absorption from an intergalactic medium which was at least 50\% neutral at $z > 6$ \citep{kashikawa2006,verhamme2015}.

\section{The Cosmic LyC Background from High-Redshift Galaxies}\label{sec:bk}

Studies of the cosmic microwave background (CMB) suggest that reionization was complete about 1 billion years after the Big Bang.  
Directly probing this time observationally is extremely challenging, although future observations of the 21 cm hyperfine transition line of neutral hydrogen, redshifted to frequencies below 200 MHz, may be able to do so. 
At present, however, only very luminous AGNs were bright enough to provide estimates of the optical depths of LyC photons \citep{prochaska2009}.  
Some constraints are provided by measurements of the cosmic microwave background wherein photons Thomson scatter 
both during and after the period of reionization, but prior to a significant reduction of the free election density due to the expansion of the Universe.  
This process acts to smooth CMB anisotropies on smaller scales, but also introduces new polarization anisotropies \citep{dore2007}. 
These two effects have allowed derivations of the electron column density from CMB observations, which give 
instantaneous reionization redshifts of $z = 7$ from the Wilkinson Microwave Anisotropy Probe ({\it WMAP}) and $7.7\pm 0.8$ 
from the {\it Planck} telescope, consistent with studies of high-redshift Ly$\alpha$-emitting galaxies \citep{kashikawa2006,kashikawa2013}.

There is significant evidence that the production of escaping LyC that led to reionization of the IGM continued on to lower redshifts. 
The escape fraction of LyC photons, $f_{esc}$, that ionize the IGM remains poorly determined because of the difficulty of directly 
detecting LyC photons at redshifts $z \ge 4$, owing to the increased abundance of Lyman Limit absorption Systems (LLS).
This greatly decreased LyC transmission through the IGM makes LyC emission very difficult to detect, even when a 
large fraction of it escapes from a $z \ge 4$ galaxy. 
Fortunately some observational information is available on the escape fraction of
LyC photons at $z\sim 3$ \citep[e.g.][]{steidel2018}.  
The technique generally used is to measure the ratio of continuum fluxes at 150 nm versus the LyC flux. 
\citet{inoue2006} estimate that the median $f_{esc}$ value rises from a few percent at $z \sim 2$ to a value of $\sim$20\% at z $\sim 4$ based on combining both direct observations of LyC photons, and also indirect estimates from the observed ionizing background. 

We therefore adopt a redshift evolution of $f_{esc}$ based on the assumption that the integrated Lyman escape fraction from all galaxies, while being consistent with observational constraints, exceeds the value required to reionize the Universe. 
We take the escape fraction for LyC photons to be 1.5\% for the redshift range $0 < z < 1$ (where it has been well constrained) but then rises with redshift to 10\% for redshifts $z = 2$ and to 20\% at $z = 3$ as was found by \citet{inoue2006}. 
We further take $f_{esc}$ to be constant at 20\% for $3 < z < 8$, which is the limit of our calculation \citep{vielfaure2020}.

Opacities measured from quasar absorption spectra \citep{shull2020}, 
and those implied by the free electron density derived from the {\it Planck} telescope, 
suggest that LyC is rapidly absorbed by the intergalactic medium (IGM)
at $z \ge 7$.
However once the IGM is fully ionized, further LyC emission from galaxies should be mostly unaffected by this process. 
Indeed studies have shown that the IGM contains a few LyC photons per hydrogen atom \citep{bolton2007}. 
We are further assuming that the rate of emission of UV photons exceeds the recombination rate (but this could also be redshift-dependent). 
Our basic interest is in determining the possible impact of these ionizing photons on the \gray opacity. Thus, for the extreme effect we assume the case where there is no absorption of LyC photons at redshifts between  $z = 0$ to $z = 7$. Realistically, Lyman Limit systems along the line of sight will influence the LyC opacity. We will discuss their impact later. The exact transition redshift to where the photons have already completely reionized the universe and thus any further LyC escaping 
from galaxies (barring recombination) represents an excess, is unknown.
We therefore also consider two possible cases, which include LyC absorption beyond redshifts of 5, or alternately, redshifts beyond 3, to 
measure the impact of these assumptions.

In addition to $f_{esc}$, we need to know the wavelength and redshift dependence for wavelengths below 91.2 nm. 
LyC photons in this region result primarily from massive star formation. 
Quasars also produce ionizing photons, however we restrict our attention in this work to photons generated in star-forming galaxies. 
The spectral properties of the LyC photons can then be taken to be those of O stars, which should generate the bulk of these photons in star-forming regions within galaxies. In keeping with our philosophy of calculating photon densities where possible from empirical data, we adopt spectral characteristics in the LyC from the Binary Population and Spectral Synthesis (BPASS) suite of binary stellar evolution models \citep{eldridge2017} because of its tested ability to reproduce the properties of observationally resolved stellar populations and in particular their colors. 
The BPASS templates cover the extreme ultraviolet (20 nm - 160 nm) spectral region produced for an instantaneous starburst population at a time 30 Myr after star formation. We adopt their model for binary evolution and a metallicity 0.0005 Z$_\odot$. 
We scale the spectra to the luminosity at 150 nm from our calculation. 

\citet{eldridge2017} provide templates that are scaled to a common luminosity at this wavelength. 
This is the highest reasonable case for LyC production, as the LyC spectrum softens with both increased metallicity and for single O stars. 
The spectrum template normalized to the luminosity density at 150 nm in our calculation, combined with our $f_{esc}$ function, determines the density of LyC photons that escape from galaxies as a function of redshift.  
We note that this procedure assumes that there are no further sources of differential extinction between 150 and 91.2 nm. In particular, we are ignoring the effects of differential dust reddening in the far-UV.  
We adopt this simplified assumption because the most primitive galaxies common at very high redshifts will have very low metallicities, which should not favor the formation of dust grains. 
Indeed, available observations show that their far-UV continuum tends to be exceptionally blue--so blue that even with the most extreme starburst models, there is very little room for any UV extinction, even at the level of a few tenths of a magnitude \citep{finkelstein2010}. 
In fact, detailed fitting of multiwavelength spectral energy distributions of $z \sim 3$ galaxies gives by far the most frequent estimate of visual extinction of $A_V = 0.0$ \citep{smith2020}.
As with other inputs to our calculation, we have chosen assumptions allowing for the maximum possible escaping flux of ionizing photons from high-redshift galaxies. 
The average effects of differential dust reddening at high redshifts are admittedly uncertain, but our assumption that it is negligible is not likely to be too far in error. 
Our results thus provide a firm upper limit to the effects of ionizing photons in the extragalactic background spectrum.

For wavelengths shorter than 22.4 nm, beyond the He II ionization edge, the spectra of even the hottest O stars cut off sharply.
For example, the ionizing spectrum of the hottest BPASS model--that is entirely binary-star-dominated and with the lowest metallicity--has a
factor of 40 drop from the red side to the blue side of the He II Lyman limit \citep{eldridge2017}. 
This very small amount of He II-ionizing photons is sufficient to explain the very weak He II emission lines seen in most green pea
galaxies \citep{saxena2020}.
Even in this most extreme case, the number of photons shortward of 22.4 nm is so small that it would contribute negligibly to the gamma ray opacity of the Universe. 

\subsection{LyC Escape from Active Galactic Nuclei}

The AGN contribution of ionizing photons is still uncertain.  Some modeling has suggested that, at intermediate redshifts ($ z \sim$ 2--3), active galactic nuclei may rival or exceed star-forming galaxies 
in the production of ionizing photons \citep{smith2020,faucher2020}. 
In one detailed calculation of the extragalactic background of ionizing photons,
quasars dominate over star-forming galaxies for lower redshifts up to $z \le 3$  \citep{khaire2019}.
The observational support for this is not entirely clear.  One result, for example, indicates that  star-forming galaxies generate three times as much ionizing flux as those from AGN in the range of z $\sim$ 2--3 \citep{steidel2018}.

Observations at $z \ge 4$ do not support a substantial AGN ionizing component \citep{parsa2018}.
Most investigators have argued that at higher redshifts, star-forming galaxies dominate over quasars in producing ionizing UV photons 
(see e.g., the discussion by \citet{puchwein2019}).
Given the uncertainties in the AGN models, we have conservatively limited our calculation to the better constrained contribution of LyC photons 
from O stars, particularly in high-ionization galaxies. 

\section{New Determination of \gray absorption at {\it Fermi} Energies}\label{sec:tau}

We use here our new determination of the flux of LyC photons in order to revise our previous calculations  of 
\gray absorption  \citep{stecker2016}, taking account of their effect on extragalactic high energy \grays via the annihilation of such \grays by $e^{+}-e^{-}$ pair production interactions. 

The threshold for photon-photon annihilation into an electron-positron pair is determined
by the Lorentz invariant quantity given by the square of the center-of-momentum energy, $s$,
where
\begin{equation}
s = 2\epsilon E_{\gamma} (1-\cos\theta) = 4m_{e}^2
\label{eqn:s}
\end{equation}
with $\epsilon = h\nu$ is the energy of the soft photon.

The absorption coefficient $\tau (E_{0},z_e)$ for a photon emitted at redshift $z_e$ where $E_{0}$ is the \gray energy as observed at redshift zero is given by
\begin{equation} 
\label{eqn:tau}
\tau(E_{0},z_{e})=c\int_{0}^{z_{e}}dz\,\frac{dt}{dz}\int_{0}^{2}
dx\,\frac{x}{2}\int_{0}^{\infty}d\nu\,(1+z)^{3}\left[\frac{u_{\nu}(z)}
{h\nu}\right]\sigma_{\gamma\gamma}(s)
\end{equation}
where $s=2E_{0}hc/\lambda x(1+z)$,  and
$u_{\nu}(z)$ is the co-moving energy density of the photon field as a function of redshift 
and $x = 1-\cos\theta$, with $\theta$ being the angle between the \gray
and the background photon \citep{stecker1992}. 
In our case this would be a LyC photon.
 
The value for $dt/dz$ in a $\Lambda$CDM Universe is given by
\begin{equation}
\frac{dt}{dz}{(z)} = {[H_{0}(1+z)\sqrt{\Omega_{\Lambda} + \Omega_{m}(1+z)^3}}]^{-1}.
\label{cosmology}
\end{equation}
The pair production cross section is given by
\begin{equation} 
\label{sigma}
\sigma_{\gamma\gamma}(s)=\frac{3}{16}\sigma_{\rm T}(1-\beta^{2})
\left[ 2\beta(\beta^{2}-2)+(3-\beta^{4})\ln\left(\frac{1+\beta}{1-\beta}
\right)\right],
\end{equation} 
\noindent where $\sigma_T$ is the Thompson scattering cross section and $\beta=(1-4m_{e}^{2}c^{4}/s)^{1/2}$. The effect of the $\beta$ dependence of the cross section is to spread out the effect of absorption to energies around the value $4m_e^2/\epsilon$.

We particularly note that it follows from equation (\ref{eqn:s}) that LyC photons effect \grays at lower energies than previously considered.

\section{Results}

\subsection{Co-moving Photon Density and the EBL}

In a series of papers \citep{stecker2012,stecker2016,scully2014}, we derived a model-independent determination of the EBL, entirely based on observationally derived luminosity functions using up-to-date local and deep galaxy survey data and color determinations. 
Those estimates covered the entire wavelength range from the far UV to the far IR, directly deriving both the $\gamma$-ray opacity and its observational uncertainties, as a function of both energy and redshift, out to a redshift of $\sim 5$ and out to TeV energies \citep{stecker2016}. 
However,  the previous work ignored the effect of LyC photons on the EBL or on \gray absorption, 
based on the previously accepted wisdom that the flux of LyC photons was negligible. 
With the recent discovery of LyC production and escape in such objects as the green pea galaxies, we have now extended our previous methods to wavelengths short-ward of the Lyman limit.

We  follow the methods of our previous work in determining, as a function of redshift, the intensity spectrum of the intergalactic background light based on multiwavelength observational data from galaxy surveys. 
Previous work quantified the observational uncertainties in the empirical determination through 68\% confidence upper and lower limits. 
For purposes of the present work, we instead derive a ``best-fit'' case in order to see more clearly the effects of the LyC and to compare them more directly with existing \gray data. 
We use previous compilations of galaxy survey data.
In cases in which they were not provided,  LDs were computed by integrating fits to the observationally determined luminosity functions. The resulting luminosity density fits and their redshift dependence mostly covered the range from the FUV to the far infrared. 
However some gaps existed particularly at higher redshifts.  
In those cases we supplemented the LDs with use of continuum colors between the wavelength bands as was done in \citep{stecker2012}.  

\begin{figure}[ht!]
\plotone{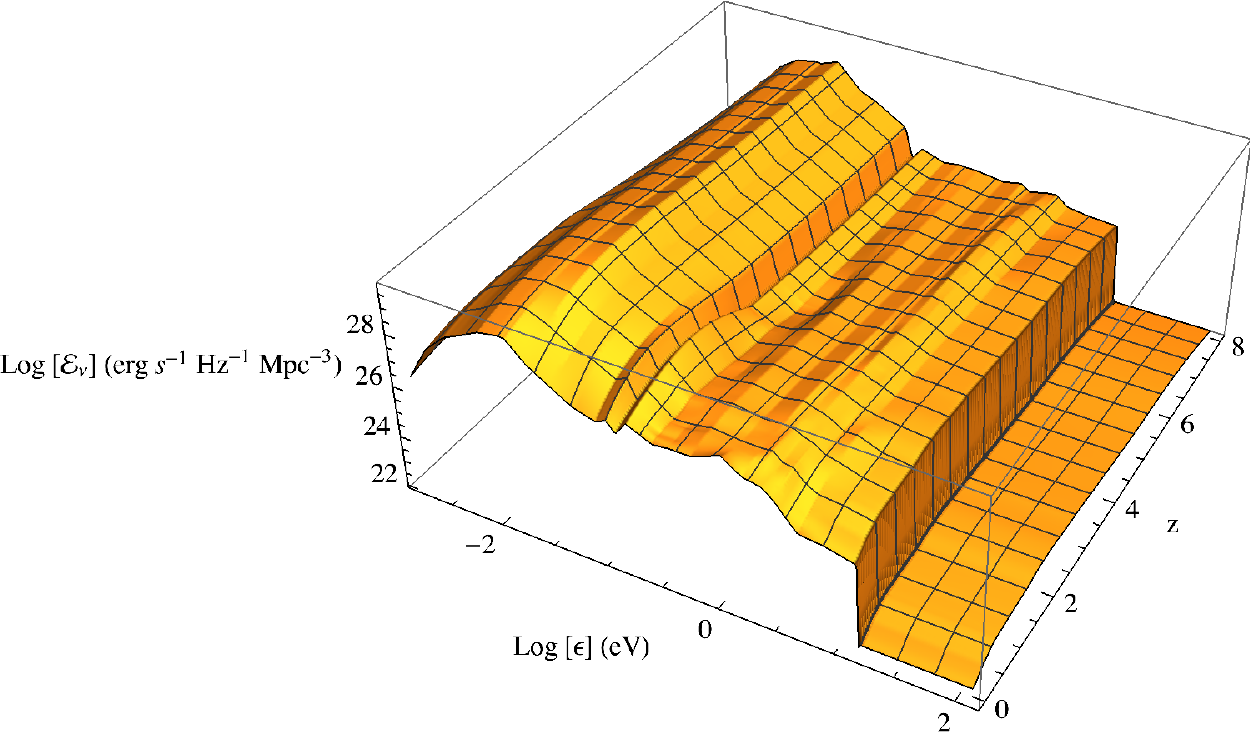}
\caption{Rest frame photon emissivities for our best-fit observationally derived case, with no LyC photons, as a function of photon energy and redshift.}
\label{fig:restframeave}
\end{figure}

\begin{figure}[ht!]
\plotone{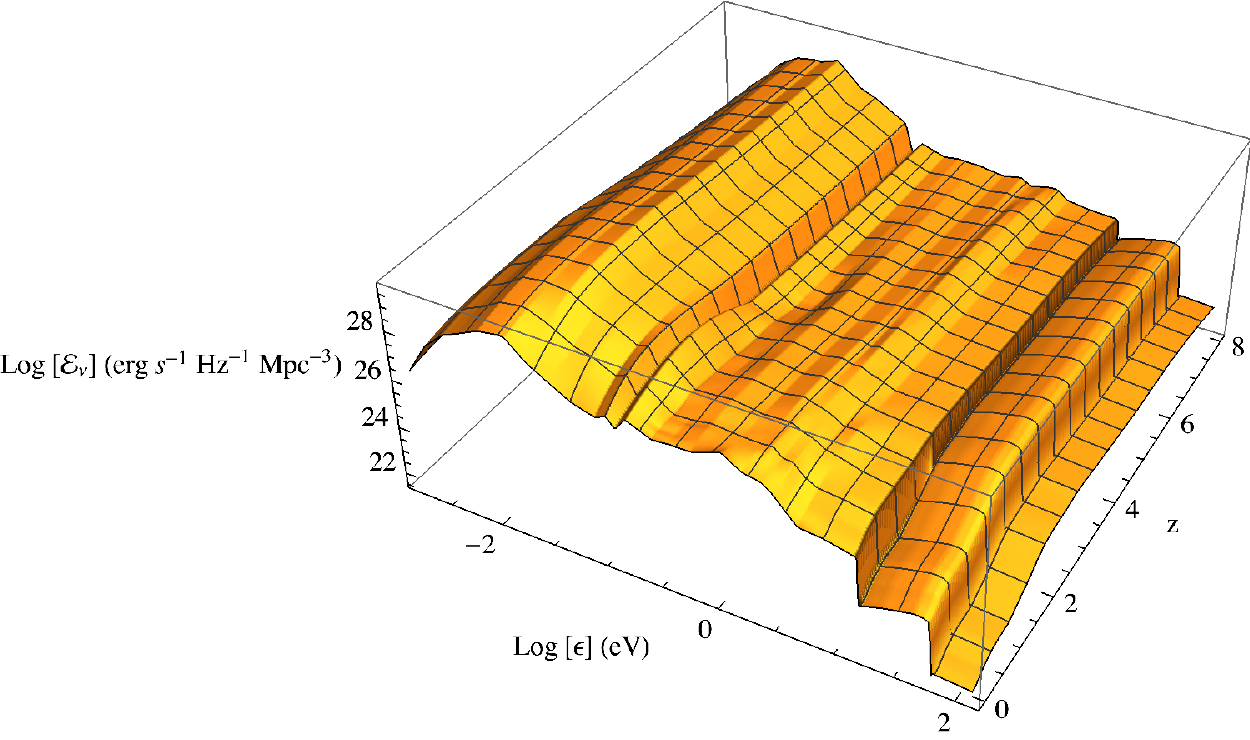}
\caption{Rest frame photon emissivities including the additional LyC component as a function of photon energy and redshift.}
\label{fig:restframelyc}
\end{figure}

The redshift evolution of each wavelength band was derived by using a rational fitting function to the redshift-luminosity density data and any necessary color-shifted data. 
This allowed us to make overall as few assumptions as possible about the luminosity density evolution. 
The chosen rational fitting function took the form of a broken power-law dependent on $(1+z)$. 
The 68\% confidence range of the fits in each waveband are the $\pm 1 \sigma$ confidence band on the luminosity density. 
Here we instead report the best fit to the data rather than the confidence band. 
We make a comparison of the two models that differ only by the inclusion or exclusion of the LyC photons that we add following the prescription of the previous section. 
Figures \ref{fig:restframeave} and \ref{fig:restframelyc} show our best fit galaxy rest frame $\rho_{L_{\nu}} = {\cal E}_\nu(z)$ excluding and including the additional LyC photons respectively.

The co-moving radiation energy density, $u_{\nu}(z)$, is
then derived from the co-moving specific emissivity ${\cal E}_{\nu}(z)$ from its time integral:
\begin{equation} 
\label{eqn:u1}
u_{\nu}(z)=
\int_{z}^{z_{\rm max}}dz^{\prime}\,{\cal E}_{\nu^{\prime}}(z^{\prime})
\frac{dt}{dz}(z^{\prime})e^{-\tau_{\rm eff}(\nu,z,z^{\prime})},
\end{equation}

\noindent where $\nu^{\prime}=\nu(1+z^{\prime})/(1+z)$ and $z_{\rm max}$ represents the
redshift corresponding to initial galaxy formation and with ${dt/dz(z)}$ given by equation (5).

\noindent We choose the values of $\Omega_{\Lambda} = 0.67$ and $\Omega_{m} = 0.33$ consistent with the latest CMB results.  
The opacity factor, $\tau_{\rm eff}$, for wavelengths longer than the Lyman limit is dominated by dust extinction. 
However, since we have used actual galaxy observations rather than  models, dust extinction is already included in ${\cal E}_{\nu}(z)$.  

We consider an opacity $\tau_{\rm eff}$ for frequencies above the rest frame Lyman limit of $\nu_{LyL} \equiv 3.29 \times 10^{15}$ Hz to that is equal to unity for the highest-background model. We shall also consider higher opacity values derived from observational constraints on HI mean free paths using 912 \AA ~data from \citet{prochaska2009} and \citet{worseck2014} at redshifts $z>3$ to see the impact of this parameter on our results.

\begin{figure}[ht!]
\plotone{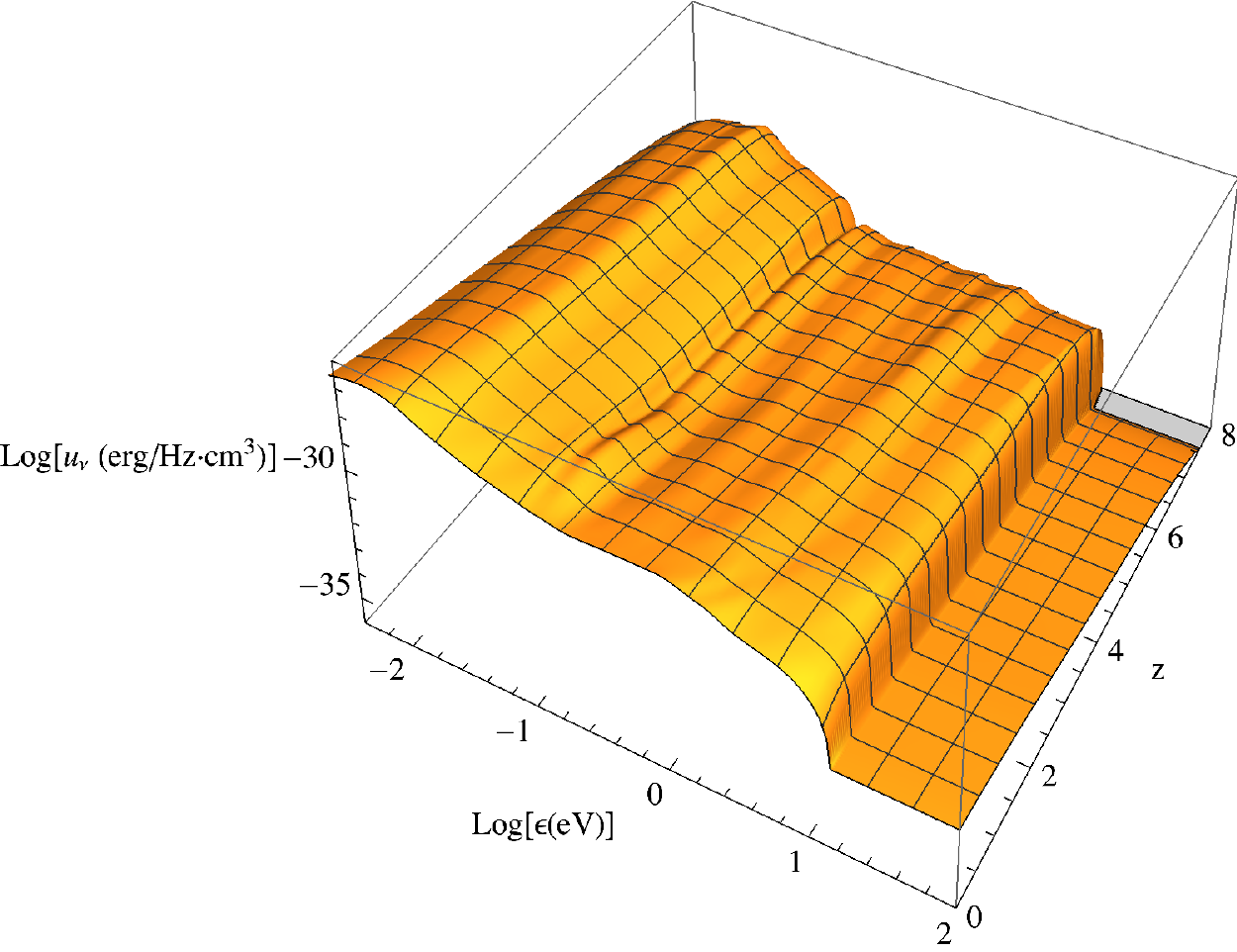}
\caption{Co-moving photon energy density derived from our best-fit case shown as a continuous function of energy and redshift, with no LyC photons.}
\label{fig:comovingave}
\end{figure}

\begin{figure}[ht!]
\plotone{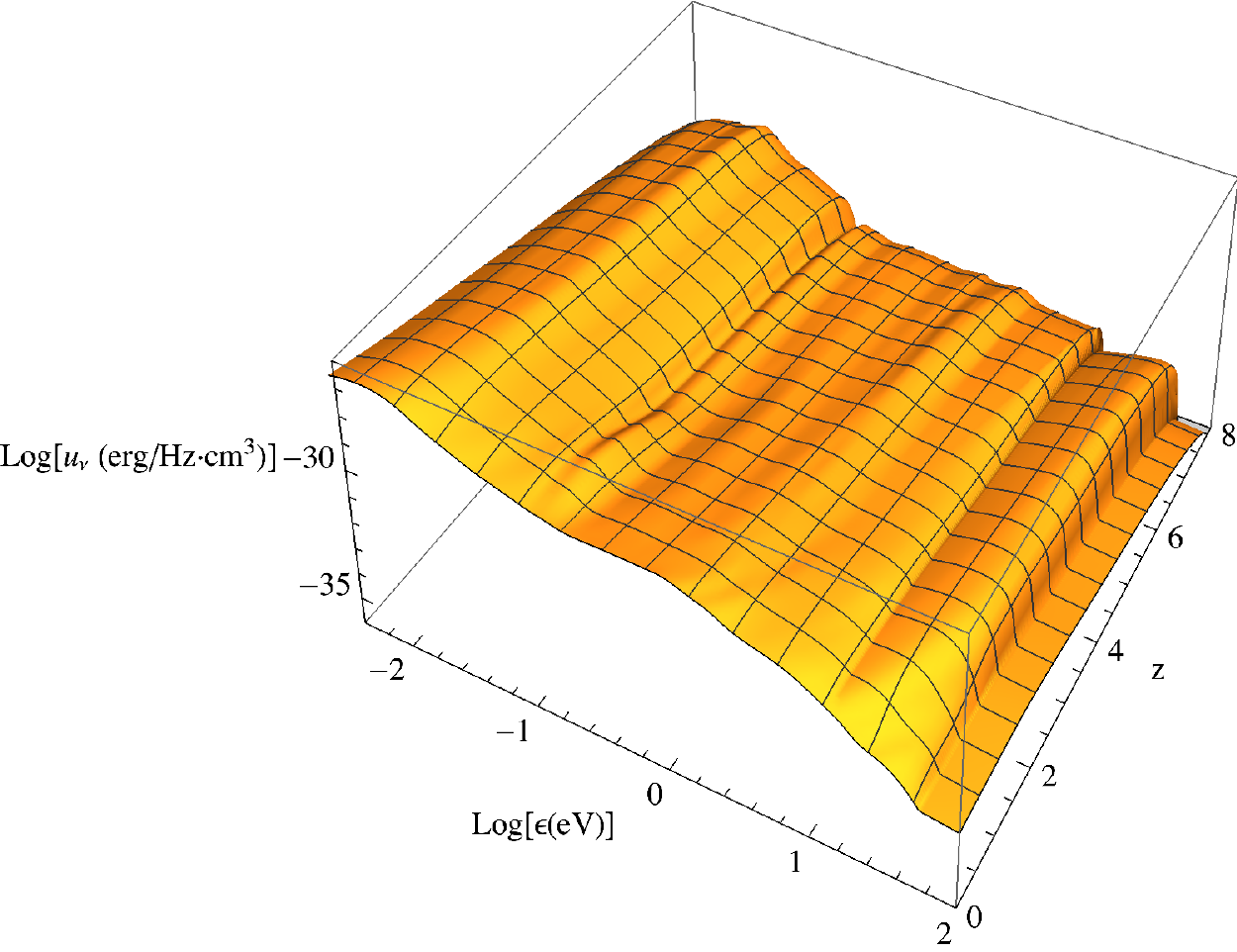}
\caption{Co-moving photon energy density derived for our best-fit case with the additional LyC component shown as a continuous function of energy and redshift.}
\label{fig:comovinglyc}
\end{figure}

Using the specific emissivity derived from the luminosity densities ${\cal E}_{\nu}(z) = \rho_{L_{\nu}}(z)$ and the inclusion of the LyC photons as 
described in section 3, the co-moving radiation density is then generated using equation \ref{eqn:u1}.  Figures \ref{fig:comovingave} and \ref{fig:comovinglyc} show the resulting co-moving photon densities, which will be further used as input in the determination of the \gray optical depths detailed in the next section.

\begin{figure}[ht!]
\plotone{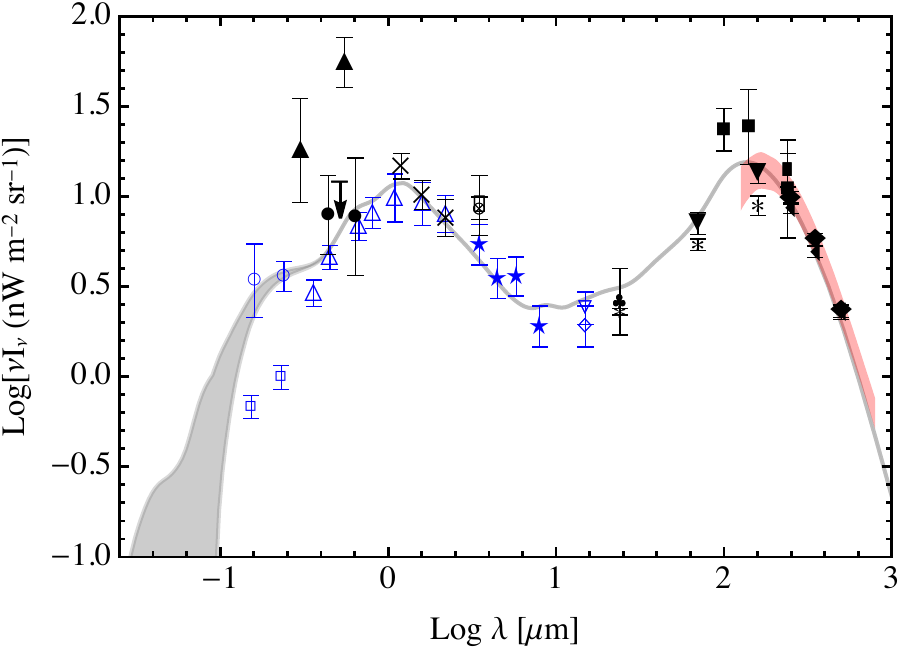}
\caption{The present extragalactic background for $z = 0$ in $nWm^{-2}sr^{-1}$ as a
function of wavelength shown as a log-log plot. The lower boundary shows the best fit to the results from \citet{stecker2016}. The upper boundary includes the contribution of the
LyC photons as calculated here. 
\citet{stecker2016} provided a listing of the EBL measurements and limits plotted and their associated references.}
\label{fig:ebltotal}
\end{figure}

We have also determined the local intergalactic background light at $z = 0$, commonly known as the extragalactic background light (EBL). 
The specific co-moving intensity of the EBL per unit solid angle, at redshift $z = 0$ is given by integrating over cosmic history. 
Using equation (\ref{eqn:u1}), together with our empirically based determinations of our specific emissivities, ${\cal E}_{\nu}(z)$, we have computed the EBL for our best-fit model both including and excluding LyC photons.  
The result is given in Figure \ref{fig:ebltotal} along with the relevant direct and indirect measurements from the literature (summarized in \citet{stecker2016}.

The grey band in Figure \ref{fig:ebltotal} is the difference between these cases with the upper edge representing the case including LyC photons. 
While present data are insufficient to distinguish between these cases, future EBL measurements may constrain the contribution of LyC photons to the EBL flux.

\subsection{Resulting Gamma-ray Opacities}

Following the prescription of section \ref{sec:tau}, we use our co-moving photon densities to determine a grid of gamma-ray opacities as a function of \gray energy spanning 1 GeV $\le E_\gamma \le$ 10 TeV and redshifts of 0 $\le z \le$ 8.  
We have calculated this grid in energy increments of 0.1 in log $E_\gamma$ and 0.1 in redshift for both our NoLC case and our (higher) LC case.  
We stress that the latter should be taken as the largest possible impact of LyC photons and may be an overestimation, 
as our intention in constructing this case was to make assumptions that would yield the maximum contribution.  
We will discuss the comparison of both cases to the present observational data in the following subsections. 
We make available this grid of opacities to the community along with our photon density calculations (both rest frame and co-moving) on-line at
\url{http://csma31.csm.jmu.edu/physics/scully/grays}.

\begin{figure}[ht!]
\plotone{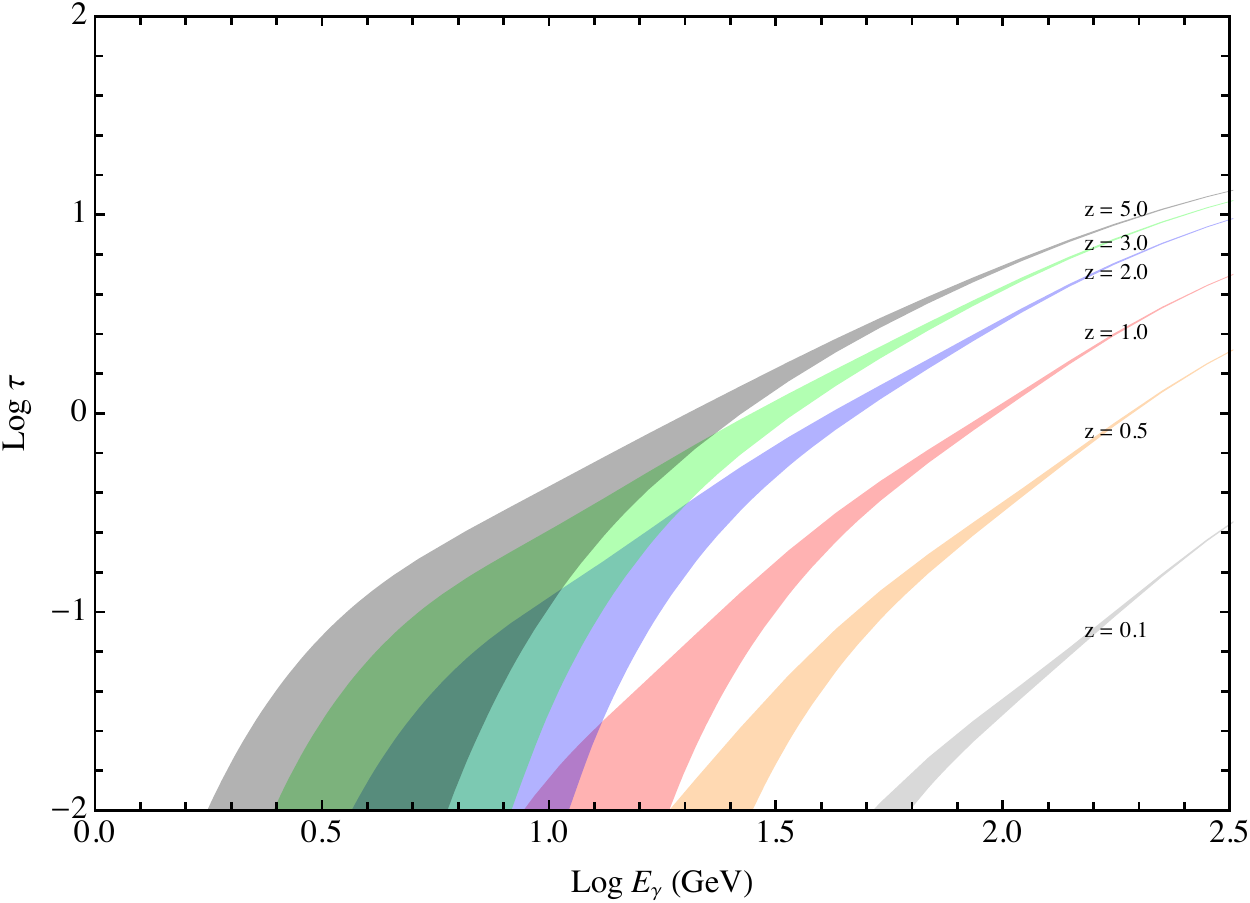}
\caption{The optical depth of the universe from the IBL as a function of energy, given for redshifts of 0.1, 0.5, 1, 2, 3, and 5. The shaded region brackets the likely contribution from LyC photons as the upper limit of the region includes our (high)  LyC case while the lower boundary is the best-fit case absent any LyC photons (NoLC). The regions narrow with increasing $E_\gamma$ as the contribution from the LyC shrinks.}
\label{fig:opactot}
\end{figure}

Figure \ref{fig:opactot} shows the resulting \gray opacities for interactions with IBL photons given for sources at z = 0.1, 0.5, 1, 2, 3 and 5 for \gray energies ranging from 1 Gev $\le E_\gamma \le$ 300 GeV representing the range with the highest impact of the additional LyC photons. 
Note that at the higher energies and redshifts our NoLC case and LC cases converge since the LyC photons can not impact the opacities in this regime. 
As expected, the greatest impact from the LyC can be seen at the higher redshifts and lowest energies where the \gray measurements required to determine the contribution from LyC exactly are challenging due to a paucity of bright \gray objects for $z \ge 3.0$. 
Nevertheless, we show that the contributions are not negligible and the \gray data even at lower redshifts hints at the possibility of determining the contribution from LyC photons and the epoch of reionization.  

\subsection{Comparison to {\em Fermi}-derived Gamma-ray Opacities}
\begin{figure}
\plotone{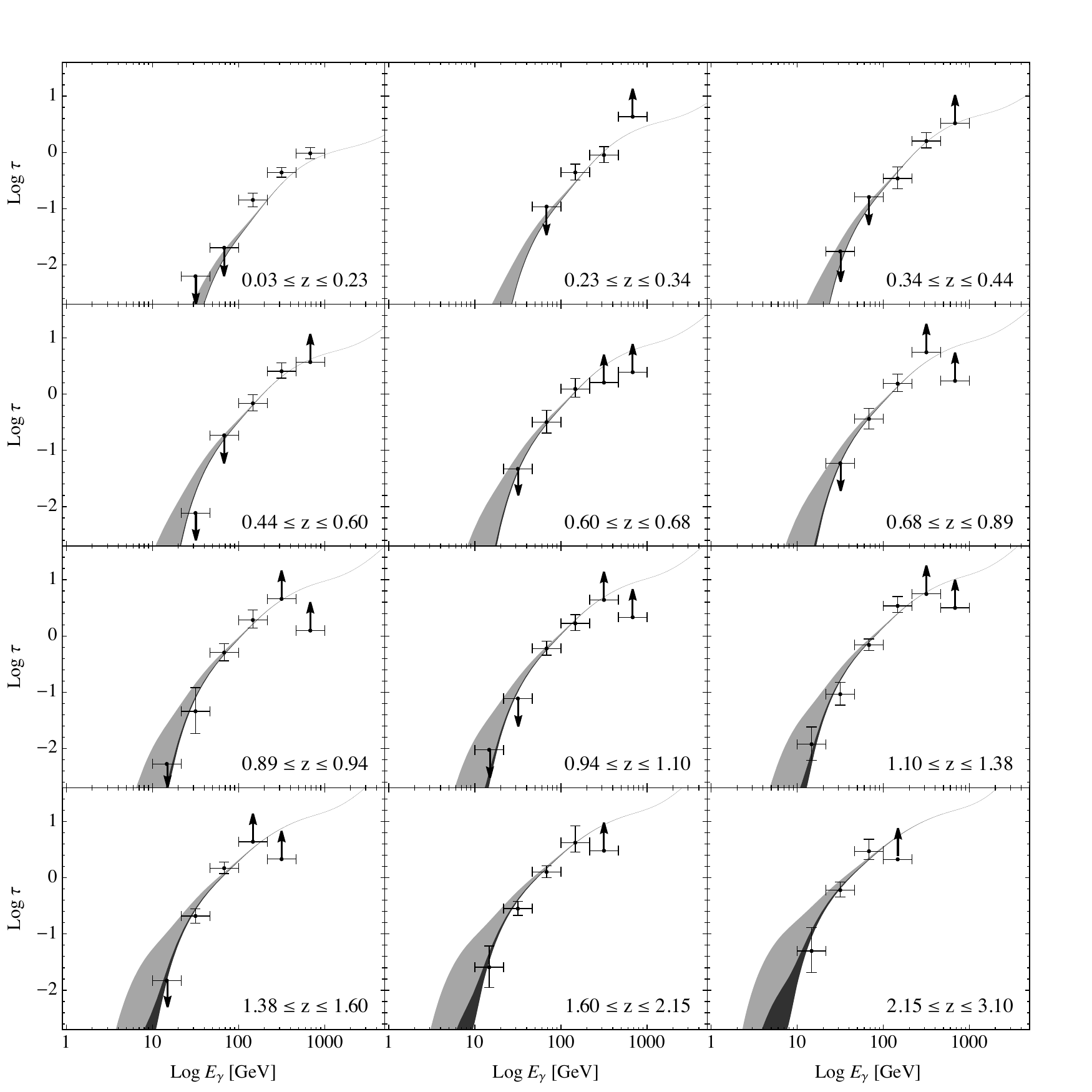}
\caption{Redshift binned optical depth measurements derived from the stacking analysis based on {\it Fermi}-LAT data of \citet{desai2019} are shown in comparison with our optical depth calculations. 
The top boundary of the grey shaded region includes LyC photons as calculated here. The top boundary of the black shaded region shows the result of using the lower escape fraction evolution assumed by \citet{khaire2019}. The bottom of the black region is the case where LyC photons are completely excluded (NoLC).} 
\label{fig:opacities}
\end{figure}

\citet{desai2019} derive \gray optical depths based on a sample of 739 blazars observed by {\it Fermi}-Large Area Telescope (LAT). 
They derive from this data set optical depth measurements utilizing a stacking analysis. 
The data set has sufficient energy and redshift coverage for them to constrain \gray opacities in 12 redshift bins covering a range of $ 0.03 \le z \le 3.1$.  
This analysis represents the most comprehensive set of \gray optical depth measurements derived strictly from \gray data produced to date.  
The {\it Fermi}-LAT derived data-set probes the UV background, including the LyC out to the high redshifts necessary for us to test the impact of the additional LyC photons.  
We plot the results of our calculation both including and excluding LyC photons against this data set in Figure \ref{fig:opacities}.

Based on the present state of the {\it Fermi}-LAT data, our (original) NoLC model excluding LyC photons is in excellent agreement with the stacking analysis in all twelve redshift bins further validating our empirical approach to determining \gray opacties based strictly on the observational data.  
Note that given the uncertainties in the stacking data, the higher LC case cannot be excluded, and moreover, despite the large numbers of blazars that were used in the analysis, the {\it Fermi}-LAT data lacks sufficient resolution to discriminate between the models, to constrain the amount of LyC photons that may be present.

In addition to our (higher) LC model, we have run cases that suppress the LyC photons for redshifts greater than 3. We have calculated an HI opacity directly from observational constraints on the mean free paths from \citet{prochaska2009} and \citet{worseck2014}.  For these data, we introduce a fit to the mean free path and directly compute the resulting LyC opacity. Note that due to observational challenges, such direct constraints only exist for $\sim$ z $>$ 3. 
At the current time it is unknown at which redshift the reionization process fully finishes, and so a LyC excess may be present in the IGM at lower redshifts.  
Our maximum LC model assumes that the ionizing photon excess begins around a redshift of $\sim 7.7$ corresponding to the instantaneous reionization redshift determined from {\it Planck}; therefore we did not include any additional LyC opacity in this model. 
The ``instantaneous reionization'' redshift can be thought of rather as a mean redshift for reionization and is likely not the end of reionization. 
Due to the analysis of the CMB optical depth by {\it Planck} and also ionization levels inferred from high-z QSOs \citep[e.g.][]{fan2006,becker2015,romano2019}, it appears that reionization was a rapid event lasting likely only for a short time period of $\Delta z \le 2.8$. 
Thus, if the IGM is not fully reionized at redshifts $z < 7$, the opacity in the neutral hydrogen will suppress a large fraction of the LyC leaking from galaxies at these times as hydrogen continues to absorb LyC photons. 

Models that include some HI opacity therefore, in principle, probe the redshift dependence of reionization. Unfortunately, even for the case where full reionization does not occur until a redshift of $\sim$ 3, where it is already clear from observations that there are more LyC photons than hydrogen atoms, there is no clearly detectable difference in predicted \gray opacities even in the highest redshift bin for which we have stacking data ($2.15 \le z < 3.1$). We have explicitly tested this using the LyC opacities derived from \citet{prochaska2009} and \citet{worseck2014} at redshifts z $>$ 3 and found the impact on the \gray opacities as presented in figure \ref{fig:opacities} to be negligible.

With direct observation of LyC photons in the early universe proving to be elusive, their indirect impact on \gray opacities may instead provide us with clues as to cosmic sources of reionization.  We have shown that the inclusion of these additional photons beyond the Lyman limit in our calculations increases \gray opacities at  energies $<$ 100 GeV and $z < 3$. We have also calculated the \gray opacity using the lower f$_{esc}$ evolution of \citet{khaire2019} that assumes far fewer LyC photons escape their source galaxies for redshifts $z < 3$. The results are shown in Figure \ref{fig:opacities}. This scenario shows little difference from the NoLC case (complete lack of LyC photons in the IGM) for $z < 1.5$ but still has a potentially measurable effect at higher redshifts and \gray energies below 10 GeV.

 
Indeed, it is evident from figure \ref{fig:opacities} that the critical test for this effect is a comparison with the measured opacity that corresponds to a \gray energy of $\sim 10~\rm{GeV}/(1+z)$. 
In this energy range the opacity is less than 0.1, leading to a high uncertainty in trying to measure it.
However, we note that additional {\it Fermi}-LAT observations in this energy range, or future \gray missions that resolve additional higher-redshift blazars, could definitively determine the contribution of extragalactic LyC photons to \gray opacities. 
This could then determine the redshift at which the end of the epoch of reionization occurred. 

\subsection{The Gamma-Ray Horizon}

\begin{figure}
\plotone{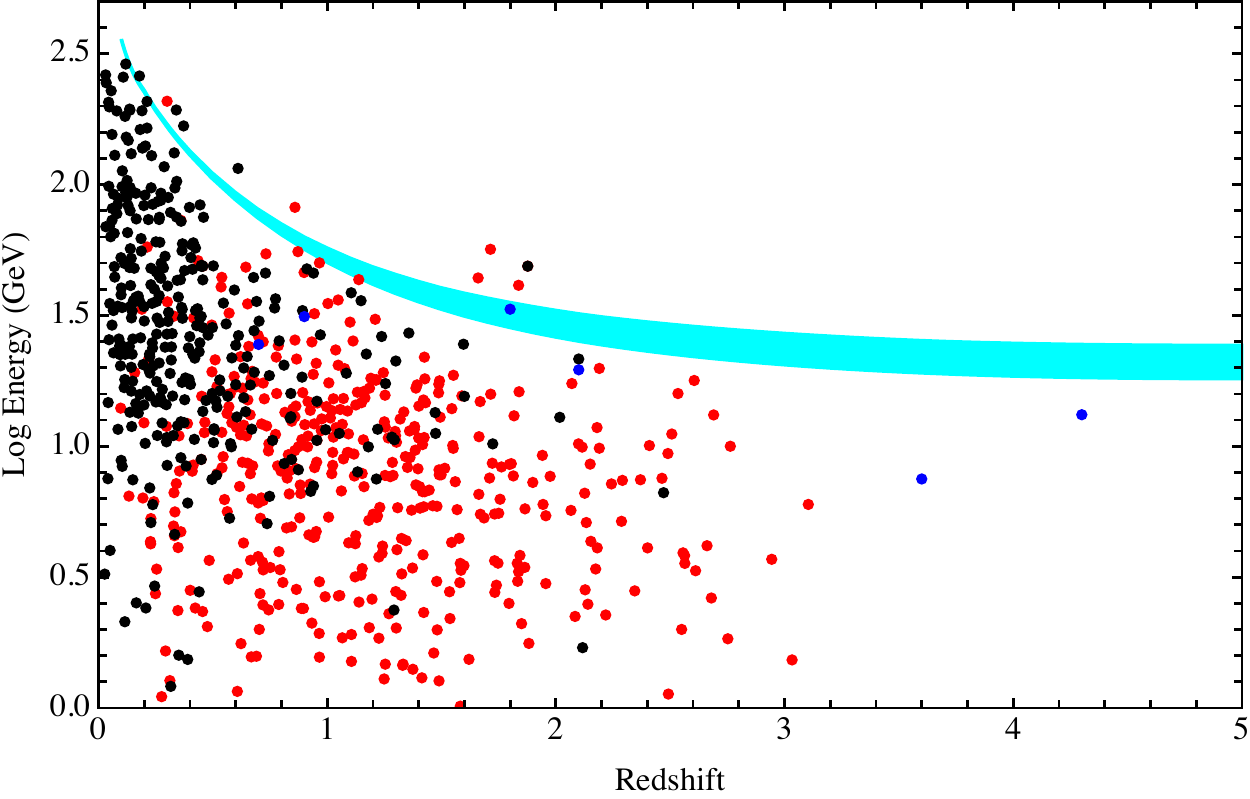}
\caption{The $\tau(E_{\gamma},z) = 1$ \gray "horizon" plot showing our results with and without including LyC photons (see text) compared with the {\it Fermi} plot of their highest energy photons from FSRQs (red), BL Lac objects (black) and and GRBs (blue) vs. redshift \citep[from][]{abdo2010,fermi2018}.}
\label{fig:horizon}
\end{figure}

The critical optical depth relation, $\tau (E,z)=1$, resulting from the annihilation of \grays by $e^{+}-e^{-}$ pair production interactions with 
extragalactic low energy photons (see equation (\ref{eqn:tau})), defines an extragalactic \gray "photosphere", or "horizon" as first shown by 
Fazio \& Stecker (1970). Beyond this horizon \grays are severely attenuated. Figure \ref{fig:horizon} shows the  critical optical depth that we 
obtained from our opacity results both with and without including the LyC photons calculated here. 
The upper limit on the opacity includes effect of the extragalactic LyC photons (LC) and the lower limit is the best fit from our previous calculated 
confidence band (NoLC) \citep{stecker2016}. Our results are shown along with the {\it Fermi} plots of the highest energy photons from sources 
as a function of redshift as given in {\it Fermi} Collaboration papers \citep{abdo2010,fermi2018}.

\subsection{Impact on Gamma-Ray Spectra}

It can be shown the general effect of interactions of $\gamma$-rays with photons originating in galaxies as discussed in Section 1 is to steepen the $\gamma$-ray spectrum at the higher energies \citep{stecker2006}.
The fourth catalog of AGNs (4FGL/4LAC) \citep{ajello2020} recently released reports on  \gray observations of 2863 high-latitude objects covering the energy range from 50 MeV to 1 TeV primarily from {\em Fermi}-LAT but also including all ground-based Cherenkov telescopes measurements of AGN. 
98\% of these objects are classified as blazars.  
The catalog contains an analysis of an eight year period of measurement from 2008 to 2016 and includes correlation with counterparts in other wavebands, blazar classifications, light curves, along with the determination of all other bulk \gray properties of the sources wherever possible. 
The sources in the 4FGL/4LAC catalog generally show significant spectral curvature. 
However, the catalogue does list an average photon index, $\Gamma$, which is useful in comparing sources' spectral hardness. 

We therefore choose to explore a simple generic power-law spectrum for a \gray source and consider sources at redshifts 1.0, 2.0, 3.0, and 5.0 to see the potential absorption characteristics of the additional LyC photons. 
We adopt a method for analyzing the spectrum based on one that two of us first used for the analysis of the H.E.S.S. observations of the source 1ES0229 + 200 \citep{stecker2008}. 
We assume an intrinsic power-law spectrum of the form $E_\gamma^{-2}$ emitted by the sources over the limited observed energy range that covers approximately a decade in energy at lower \gray energies where the LyC absorption (if present) has its greatest impact on the resulting curve. 
We multiply the power-law curves by an absorption factor $e^{-\tau(E_\gamma,z)}$ choosing the appropriate redshift opacities. 
The optical depths, $\tau(E_\gamma,z)$, are determined as described in section \ref{fig:opacities}. 
We do not consider the optical depths from other models in the literature as our goal here is not to test our best-fit or LyC cases against existing models, but rather to illustrate qualitatively their impact.

\begin{figure}
\plotone{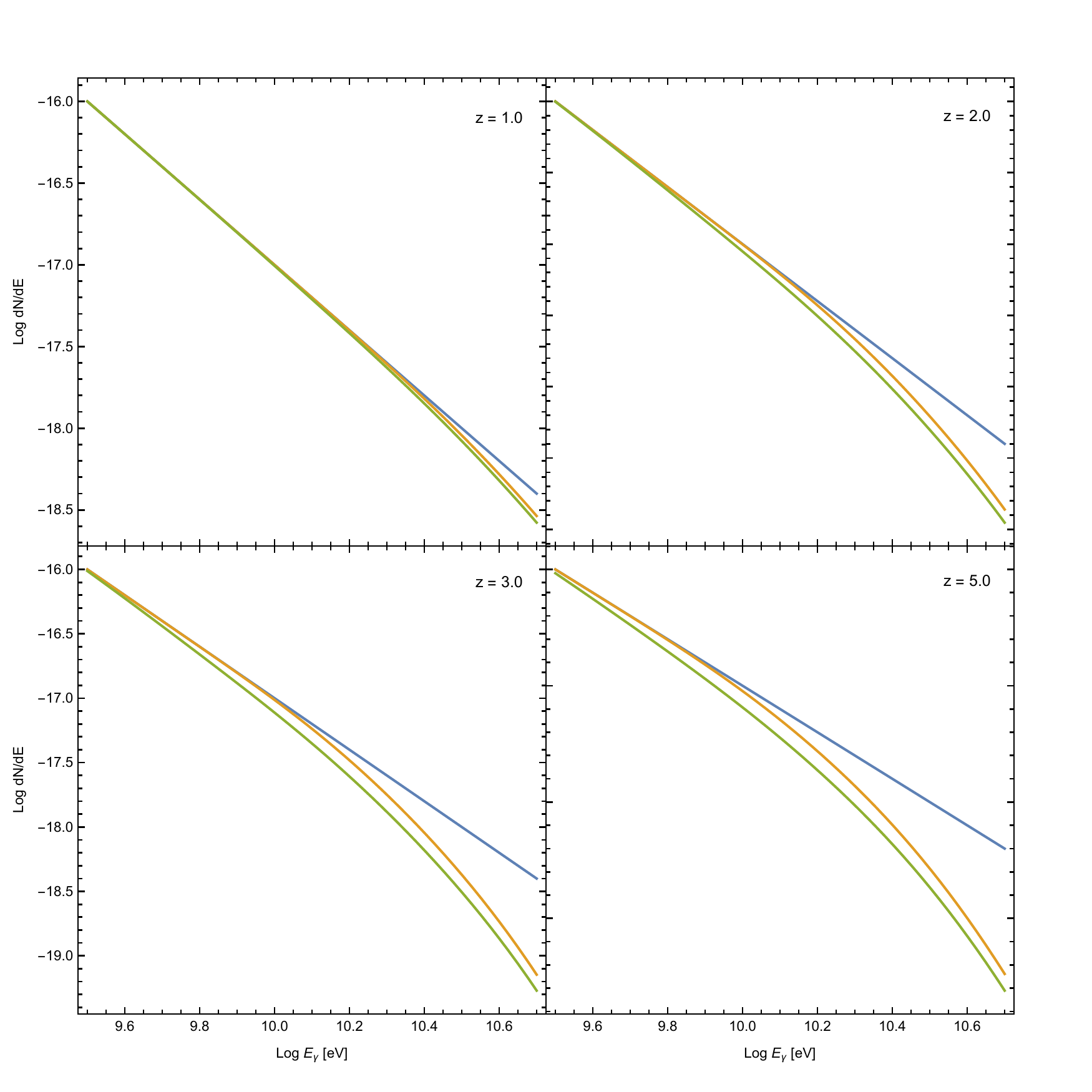}
\caption{An intrinsic power-law spectrum of $\sim E^{-2}$ (blue) adjusted for \gray absorption for our best-fit no LyC (orange) and instantaneous reionization (green) cases for redshifts of $z=$ 1.0, 2.0,3.0, and 5.0}
\label{fig:speccomp}
\end{figure}

Figure \ref{fig:speccomp} shows a comparison of our best-fit and instantaneous reionization cases as applied to the simple $E_\gamma^{-2}$ intrinsic spectrum for the selected redshifts.  
Starting with the z $\sim 3$ case, enough divergence between the cases appears to be able to potentially discriminate between the two scenarios. 
With the inclusion of LyC photons, the steepening in the intrinsic gamma-ray spectrum from absorption comes in at a lower $\gamma$-ray energy than for the NLC case with no LyC photons included, as follows from equation (\ref{eqn:s}), thus shifting the location of the energy where the \gray absorption spectrum departs from the intrinsic power-law spectrum. 
Generally speaking, additional LyC photons have some impact on \gray energies even as low as a few GeV at higher redshifts. 
The overall effect is a gradual departure from an intrinsic source spectrum.
Figure \ref{fig:speccomp} shows a comparison of our "NLC" and LC (assuming instantaneous reionization) cases, as applied to our simple $E_\gamma^{-2}$ intrinsic spectrum for the selected redshifts.  

The greater opacities that result at lower energies, as exhibited by our instantaneous reionization model, suggest that their photon indices and indeed those of other high-redshift {\it Fermi} detections would be steeper. 
This is clear from figure \ref{fig:speccomp} for the $z = 3$ and $z = 5$ cases.  
In the z $\sim 3$ case there is enough divergence between these cases may potentially allow observational discrimination between the two scenarios. 
At that redshift, the steepening in the spectrum due to absorption occurs at an energy that is a factor of three lower for the instantaneous reionization LyC case as compared to "no LyC" spectrum. 

The difference in spectral indexes between the two cases is a clear signature of the additional LyC photons. 
To illustrate this, we have computed for redshifts spanning $2 \le z \le 5$ the quantity $\Delta\Gamma$, which is the difference between the intrinsic spectral index and that generated by fitting the low-energy portion of the \gray spectrum including absorption. $\Delta\Gamma$ ranges from 0.1 ($z=2$) to 0.25 ($z=5$). 
If further high-z detections by {\it Fermi} or future \gray missions continue to show this trend, that would be strong evidence in favor of the additional contribution of LyC photons at these redshifts.

Only a handful of bright blazars are currently known at sufficiently high redshifts to explore the contribution of LyC photons to the extrinsic absorption of their spectra. 
At present, their spectra are not sufficiently determined to directly compare cases that include or exclude additional LyC photons. 
Moreover, many of the highest redshift detections occur during flaring events wherein the characteristic spectra may not be representative making it difficult to disentangle the various potential effects. 
For example \citet{paliya2019} explored the multiwavelength light curve of the 2018 flaring event of blazar DA 193 ($z\sim 2.4$).  
They found that their one-zone model fit that successfully reproduces the data over most wavelength bands leads to harder \gray spectra that they attribute to variability of the underlying electron population.
   
Despite these limitations, five high-redshift blazars $z > 3$ recently detected by {\it Fermi} \citep{ackermann2017} do have $\Gamma$ determinations. 
These five blazars are observed to have 
steeper spectra then the general trend in the 4FGL/4LAC catalog
for a steepening with redshift and luminosity. 
It is clear from figure \ref{fig:speccomp} that the greater opacities that result at lower energies, as exhibited by our instantaneous reionization model, imply that photon indices of {\em all} high-redshift \gray detections should be steeper, and is a  signature of the additional LyC photons. 
Further high-z detections by {\em Fermi} or a future \gray mission should definitively resolve the issue.

\section{Discussion}

The newly discovered green pea class of extreme emission line starburst galaxies (EELGs), and other galaxies with large escape fractions of FUV and EUV photons in the Lyman continuum (LyC) region, are the most typical galaxies found at high redshifts. 
Recent observations indicate that many more LyC photons at high redshifts escape from such galaxies into intergalactic space than was previously suspected. 
They may be the major cause of reionization of the intergalactic medium. 
The upcoming deployment of the James Webb Space Telescope will enable improved observations of green pea galaxies and extreme emission line galaxies at high redshifts, leading to an improved understanding of the production of LyC photons at high redshifts which reionized the Universe.

We have estimated the contribution of these hitherto unknown FUV and EUV photons from  EELGs to the EBL and their subsequent effect on extragalactic high energy \grays via the annihilation of such \grays by $e^{+}-e^{-}$ pair production interactions with LyC photons.
At these energies, the optical EBL optical depth of distant sources, although less than unity, increases significantly from the contribution by 
escaping LyC photons. This may have observable consequences in \gray observations of blazars.
We have compared the resulting \gray opacity predictions from EBL models with no LyC photons, and our new models with maximum
realistic contributions from LyC photons, EUV absorption from intergalactic HI, and with more restrictive possible escape fractions (see Figure 
\ref{fig:opacities}). The nature of the sources of cosmic reionization may be probed using \gray absorption measurements.
The \gray predictions could also be used to test the HI opacity of the IGM,
particularly at low redshifts, where it is very uncertain. 
The best current 
observational data from {\it Fermi} on the \gray horizon and on the average \gray optical depths based on stacking of blazar spectra 
cannot discriminate between the cases considered here.
However, all of our models appear to be accurate enough to make reasonable estimates
of \gray opacity as a function of energy and redshift. 
Thus, with foreseeable improvements in the {\it Fermi} database, it should be possible to make a determination of these effects.
In particular, in the range of several to tens of GeV, the spectra of distant blazars at $z \ge 3$ would be measurably steepened by the additional 
opacity provided by the background of escaping LyC photons.

\acknowledgments

The authors acknowledge the support of NASA through the {\it Fermi} Guest Investigator Program (Cycle 10)  Number 80NSSC17K0758. 
We thank an anonymous referee for constructive criticisms which greatly improved the manuscript.
We are grateful to Marco Ajello for generously providing both the highest energy photon \gray data and GeV-TeV \gray opacity constraints.

\end{document}